\documentclass[letterpaper, 10 pt, conference]{ieeeconf}    

\IEEEoverridecommandlockouts                             

\usepackage{mathptmx} 
\usepackage{times} 
\usepackage{cite}
\usepackage{amsmath,amssymb,amsfonts}
\usepackage{algorithmic}

\usepackage{graphicx}
\graphicspath{{./images/}}
\DeclareGraphicsExtensions{.pdf,.png}
\usepackage{wrapfig}

\usepackage{textcomp}
\def\BibTeX{{\rm B\kern-.05em{\sc i\kern-.025em b}\kern-.08em
    T\kern-.1667em\lower.7ex\hbox{E}\kern-.125emX}}
\usepackage{mathtools}
\usepackage{bbm}
\usepackage{thmtools}
\usepackage{flushend}
\usepackage{tikz}
\usetikzlibrary{positioning}
\usepackage{color}
\usepackage{dsfont}

\usepackage[style=base]{caption}
\usepackage{subcaption}

\newcommand{\dd}   {{\rm d}\hbox{\hskip 0.5pt}}

\newcommand{\rline}{{\mathbb R}}

\newcommand{\nline}{{\mathbb N}}

\newcommand{\bbm}[1]{\left[\begin{matrix} #1 \end{matrix}\right]}

\newcommand{\rfb}[1]{\mbox{\rm
   (\ref{#1})}\ifx\undefined\stillediting\else:\fbox{$#1$}\fi}

\renewenvironment{proof}{\vspace{.1cm}\noindent{\sc
    Proof.}\hspace{0.10cm}\,\,}{$\hfill\Box$\vspace{.3cm}} 

\newtheorem{theorem}{Theorem}[section]

\newtheorem{lemma}[theorem]{Lemma} 
\newtheorem{proposition}[theorem]{Proposition}

\title{\LARGE \bf Secure Formation Control via Edge Computing Enabled by Fully Homomorphic Encryption and Mixed Uniform-Logarithmic Quantization}

\author{Matteo Marcantoni$^{1}$, Bayu Jayawardhana$^{2}$, Mariano Perez Chaher$^{2}$, Kerstin Bunte$^{1}$
\thanks{This work was supported by the Dutch Research Council (NWO) under Smart Industry programme, SMART-AGENTS project No. 18024.}
\thanks{$^{1}$Matteo Marcantoni and Kerstin Bunte are with the Bernoulli Institute, Faculty of Science and Engineering, University of Groningen, 9747AG Groningen, The Netherlands (email: m.marcantoni@rug.nl; kerstin.bunte@gmail.com).}
\thanks{$^{2}$Bayu Jayawardhana is and Mariano Perez Chaher was
with the Engineering and Technology Institute Groningen, Faculty of Science and Engineering, University of Groningen, 9747AG Groningen, The Netherlands (email: b.jayawardhana@rug.nl; marianopablo97@outlook.com).}
}

\begin{document}

\maketitle
\thispagestyle{empty}
\pagestyle{empty}

\begin{abstract}

Recent developments in communication technologies, such as 5G, together with innovative computing paradigms, such as edge computing, provide further possibilities for the implementation of real-time networked control systems. 
However, privacy and cyber-security concerns arise when sharing private data between sensors, agents and a third-party computing facility.
In this paper, a secure version of the distributed formation control is presented, analyzed and simulated, where gradient-based formation control law is implemented in the edge, with sensor and actuator information being secured by fully homomorphic encryption method based on learning with error (FHE-LWE) combined with a proposed mixed uniform-logarithmic quantizer (MULQ). 
The novel quantizer is shown to be suitable for realizing secure control systems with FHE-LWE where the critical real-time information can be quantized into a prescribed bounded space of plaintext while satisfying a sector bound condition whose lower and upper-bound can be made sufficiently close to an identity.
An absolute stability analysis is presented, that shows the asymptotic stability of the closed-loop secure control system. 
\end{abstract}

\section{Introduction}
\label{intro}

Recent advances in communications technology, such as 5G, offer ultra-low latency and highly reliable wireless information exchange paving the way for pervasive edge computing for industrial internet-of-things (IoT) applications \cite{narayanan2020}. 
In fact, the combination of 5G and high-performance computing infrastructure can enable the deployment of real-time networked control systems via edge servers, where sensing and control information are exchanged in real-time within the wireless network. 

With the emergence of networked control systems, privacy and cyber-security concerns become increasingly important and cutting-edge solutions are highly desired \cite{narayanan2020,giraldo2019survey,zhang2018data,yang2017survey}, especially when using potentially untrustworthy third-party computing facilities. 
Traditional encryption algorithms, such as AES and RSA, can protect private data during their transportation to edge or cloud computing devices. 
However, the corresponding decryption process removes their confidentiality in order to allow for further data processing in these third-party servers \cite{zhang2018data}. 
Fully homomorphic encryption (FHE) algorithms have recently been introduced to overcome this shortcoming as they allow for data manipulation directly on the encrypted data, thus eliminating the need for decryption \cite{kogiso2015cyber,farokhi2016secure,kim2016encrypting}. 
Specifically, FHE based on Learning with Error (FHE-LWE) allows us to perform addition and multiplication with the encrypted data without error accumulation  due to the encryption process, thereby enabling secure control system design \cite{kim2016encrypting,kim2022dynamic,cheon2018need,Kim2020}. 

When dealing with real-world actuator and sensor data, however, a quantization process is required before FHE-LWE can be used \cite{Kim2020,chaher2021homomorphic}.
We propose the use of the mixed uniform-logarithmic quantizer (MULQ), derived from the quasi-logarithmic
quantizer \cite{chaher2021homomorphic}, because it permits both to quantize data in a prescribed bounded integer set and to adjust its lower and upper-bound, as we discuss later. In the literature, several works deal with control design and analysis methods in presence of quantization \cite{almuzakki2020nearest, persis2012, fu2005sector}. Here we focus on the sector bound approach \cite{almuzakki2020nearest,fu2005sector} to assess the absolute stability of the secure feedback control system.

In this paper, we present a secure version of the distributed formation control problem \cite{Marina-book,de2016distributed,Marina2015controlling,Oh2015survey,AnYuFiHe08}, in which the objective is to guide a multi-agent system towards a desired formation while guaranteeing privacy and cyber-security via FHE-LWE and MULQ. 
As an example, this framework could be useful when a third-party computing facility is employed to facilitate the usage of information coming from external sensors, i.e. not located on-board the agents, that are essential to the control task.
Our contribution is three-fold: we describe the distance-based secure edge control system, we discuss the sector bound property of the mixed uniform-logarithmic quantizer and we analyze the stability of the closed-loop multi-agent system. 

We organize the paper as follows. Preliminaries on graph and on formation control are presented in Section \ref{sec:preliminaries}. 
The discussion on MULQ is given in Section \ref{sec:MULQ}. The main result on secure formation control with MULQ is presented in Section \ref{sec:formation_cloud_control}. 
A simulation result on secure formation control of four agents forming a square is given in Section \ref{sec:numerical_sim}. Finally, the conclusions are presented in  Section \ref{conclusions}.

\section{Preliminaries}
\label{sec:preliminaries}

\subsection{Fully Homomorphic Encryption by Learning with Error}
\label{subsec:FHE}

Prior to describing the encryption method, we need to introduce some relevant notations as used in the literature.
Throughout this work, we denote the base 10 logarithm of $x$ as $\log(x)$.
Let the plaintext space, namely the unencrypted information space, be a bounded integer set $[a]:=\big\{b\in\mathbb{Z}:-\frac{a}{2}\leq b<\frac{a}{2}\big\}$ of cardinality $a\in\mathbb{N}$.
Since the plaintext for our control application represent numbers of sensor and control signals we can conveniently set $a$ to be a power of $10$.
Furthermore, let the cyphertext space, i.e. the encrypted information space, be the set of integers modulo $\mathbb{Z}_{q}$ whose elements are denoted in bold, $q\in\mathbb{N}$.
We define $q=wa$ where the parameter $w$ is also a power of $10$. 

For the implementation of a secure formation control law via third-party computing facility, full capability of multiplication and addition operations in cyphertext are desirable. 
This can be achieved with Fully Homomorphic Encryption by Learning with Error (FHE-LWE), briefly reviewed as follows.
Let $m\in[a]^n$ be a plaintext message of length $n$ to be encrypted and $\mathbf{s}\in\mathbb{Z}_q^N$ be a secret key of length $N$.
The encryption operation $\text{Enc}(\cdot)$ of $m$ is defined by
\begin{equation}
    \label{eq:Enc}
    \mathbf{M}^{\text{Enc}}=\text{Enc}({m}):=[(-\mathbf{A}\mathbf{s}+wm+\mathbf{e})\pmod{q},\mathbf{A}] 
\end{equation}
where the matrix $\mathbf{M}^{\text{Enc}}\in\mathbb{Z}_q^{n\times N+1}$ is the resulting cyphertext message.
Furthermore, the operators $[\cdot,\cdot]$ and $\pmod{q}$ refer to the concatenation and modulo $q$ operator respectively.
The matrix $\mathbf{A}\in\mathbb{Z}_q^{n\times N}$ is sampled from a uniform distribution over $\mathbb{Z}_q^{n\times N}$, while the {\it injected error} vector $\mathbf{e}$ is sampled from a uniform distribution over $[r]^n$ with $r<w$, such that the following inequality holds: $|\mathbf{e}_i|<\frac{w}{2}$ for $i=1,\cdots ,n$.

Once the encrypted message $\mathbf{M}^{\text{Enc}}$ is computed, the decryption process will involve the secret key vector $\mathbf{s}$ and the injected error vector $\mathbf{e}$. 
The latter one has been shown to be crucial for the security of the encryption \cite{Kim2020}.  
Let $\mathbf{M}^{\text{Enc}}\in\mathbb{Z}_q^{n\times N+1}$ be a cyphertext message and $\mathbf{\bar{s}}:=\text{col}(1,\mathbf{s})$ be a stacked column vector of length $N+1$ constructed by the secret key vector $\mathbf{s}$. 
The decryption process $\text{Dec}(\cdot)$ of $\mathbf{M}^{\text{Enc}}$ is defined by  
\begin{equation}
\label{eq:Dec}
{m}=\text{Dec}(\mathbf{M}^{\text{Enc}}):=
\left\lceil\frac{(\mathbf{M}^{\text{Enc}}\mathbf{\bar{s}})\pmod{q}}{w} \right\rfloor=\left\lceil\frac{w{m}+\mathbf{e}}{w} \right\rfloor
\end{equation}
where vector $m\in[a]^n$ is the resulting plaintext message and $\left \lceil \cdot \right \rfloor$ refers to the element-wise round half away from zero\footnote{For example, $\lceil -1.5 \rfloor=-2$ and $\left \lceil -1.4 \right \rfloor=-1$.}.

As briefly mentioned before, the FHE-LWE method allows for the addition and multiplication operations to take place in the cyphertext space. 
Indeed, let $\mathbf{M}^{\text{Enc}}_1,\mathbf{M}^{\text{Enc}}_2\in\mathbb{Z}_q^{n\times N+1}$ be two messages in cyphertext computed via \eqref{eq:Enc} and $m_1,m_2\in[a]^n$ their decrypted {\it vectors} in plaintext. One can compute that 
\begin{equation}
    \label{eq:HE+}
    \text{Dec}(\mathbf{M}^{\text{Enc}}_1+\mathbf{M}^{\text{Enc}}_2)={m}_1+{m}_2
\end{equation}
as long as $m_1+m_2\in [a]^n$ and $\big|\mathbf{e}_{1,i}+\mathbf{e}_{2,i}\big|<\frac{w}{2}$ holds for $i=1.\cdots,n$ \cite{Kim2020}.

Before defining  the multiplication operation in the cyphertext space, we need to introduce another encryption method.
Let $m_1\in [a]$ be a {\it scalar} message to be encrypted. The second encryption method $\text{Enc2}(\cdot)$ is define by
\begin{equation}
    \label{eq:Enc2}
    \mathbf{M}^{\text{Enc2}}_1=\text{Enc2}(m_1):=m_1R+\mathbf{O}^{\text{Enc}} 
\end{equation}
where the matrix $\mathbf{M}^{\text{Enc2}}_1\in\ \mathbb{Z}_q^{\log(q)(N+1)\times N+1}$ is the resulting cyphertext. 
$\mathbf{O}^{\text{Enc}}=\text{Enc}({0}_{\log(q)(N+1)})$ refers to the encrypted zero column vector ${0}_{\log(q)(N+1)}\in [a]^{\log(q)(N+1)}$ and $R$ is defined by:
\begin{equation*}
    R:=\text{col}\big(10^{0},10^{1},\cdots,10^{\log(q)-1}\big)\otimes \ I_{N+1}  
\end{equation*}
where $\otimes$ denotes the Kronecker product and $I_{N+1}$ the identity matrix of dimension $N+1$. 

Since any row vector $\mathbf{c}\in \mathbb{Z}_q^{N+1}$ can be represented as $\mathbf{c}=\sum_{i=0}^{\log(q)-1}10^{i}\mathbf{c}_i$ with row vectors $\mathbf{c}_i\in \mathbb{Z}_q^{\log(q)(N+1)}$, whose components are one of the single digit from $0$ to $9$, we can define a function $D:\mathbb{Z}_q^{N+1}\rightarrow \mathbb{Z}_q^{\log(q)(N+1)}$ that decomposes any row vector $\mathbf{c}\in \mathbb{Z}_q^{N+1}$ by its string of digits as
\begin{equation*}
    \label{eq:Decomp}
    D(\mathbf{c}):=[\mathbf{c}_0,\mathbf{c}_1,\cdots,\mathbf{c}_{\log(q)-1}]
\end{equation*}
Therefore, $\mathbf{c}=D(\mathbf{c})R$ for any row vector $\mathbf{c}\in\mathbb{Z}_q^{N+1}$ \cite{Kim2020}.

Now we have the necessary means to define multiplication in the cyphertext space.
Let $\mathbf{M}^{\text{Enc2}}_1\in\mathbb{Z}_q^{\log(q)(N+1)\times N+1}$ and $\mathbf{m}^{\text{Enc}}_2\in\mathbb{Z}_q^{N+1}$ be two messages in cyphertext computed via \eqref{eq:Enc} and \eqref{eq:Enc2} respectively and $m_1,m_2\in [a]$ their decrypted {\it scalars} plaintext.
The multiplication operation $\circledast$ of $\mathbf{M}^{\text{Enc2}}_1$ and $\mathbf{m}^{\text{Enc}}_2$ is defined as follows 
\begin{equation}
    \label{eq:mult1}
    \mathbf{M}^{\text{Enc2}}_1\circledast \mathbf{m}^{\text{Enc}}_2:=D(\mathbf{m}^{\text{Enc}}_2)\mathbf{M}^{\text{Enc2}}_1
\end{equation}
where $\mathbf{M}^{\text{Enc2}}_1\circledast \mathbf{m}^{\text{Enc}}_2\in\mathbb{Z}_q^{N+1}$, which is a row vector. 

It can be shown now that 
\begin{equation}
    \label{eq:mult2}
    \text{Dec}(\mathbf{M}^{\text{Enc2}}_1\circledast \mathbf{m}^{\text{Enc}}_2)=m_1m_2
\end{equation}
as long as $m_1m_2\in [a]$ and $\bigg|\frac{m_1\mathbf{e}_2}{w}+\frac{D(\mathbf{m}^{\text{Enc}}_2)\mathbf{e}_{1}}{w}\bigg|<\frac{1}{2}$ holds \cite{Kim2020}.
The injected error vector $\mathbf{e}_{1}\in [r]^{\log(q)(N+1)}$ comes from the encryption of the zero vector in \eqref{eq:Enc2}.
Note that $D(\mathbf{m}^{\text{Enc}}_2)\mathbf{e}_{1}$ is a scalar as $D(\mathbf{m}^{\text{Enc}}_2)\in\mathbb{Z}_q^{\log(q)(N+1)}$ is a row vector. 

\subsection{Formation graph with infinitesimal rigid formation and distance-based formation control}
\label{subsec:formation}

Following the rigidity formation framework as proposed in \cite{AnYuFiHe08}, we will define the formation control using an undirected graph $\mathbb{G}=(\mathcal{V},\mathcal{E})$ where $\mathcal{V}=\{1,2,\ldots,n\}$ is the set of vertices, representing the set of $n$ mobile agent, and $\mathcal{E}\subseteq \mathcal{V}\times \mathcal{V}$ is the set of edges. 
Note that for each pair of agents $i$ and $j$ we define only one edge in $\mathcal E$, i.e. either $(i,j)$ or $(j,i)$. 
For the pair $(i,j)$ the agent $i$ is referred to as the tail node and the agent $j$ as the head node.  
The set of neighboring agents to the $i$-th agent is denoted by $\mathcal{N}_i:=\{j\in\mathcal{V}: (i,j)\in\mathcal{E}\vee (j,i)\in\mathcal{E}\}$. 

Associated to the graph $\mathbb G$, we can define an incidence matrix ${B}\in\mathbb{R}^{|\mathcal{V}|\times |\mathcal{E}|}$ whose elements $b_{ik}$ are given by
\begin{equation}
    \label{eq:Bincidence}
    b_{ik}=\left\{\begin{matrix}
                +1\ \text{if}\ i=\mathcal{E}_k^{\text{tail}}\\ 
                \ -1\ \text{if}\ i=\mathcal{E}_k^{\text{head}}\\ 
                \ 0 \ \text{otherwise}\enspace,
\end{matrix}\right.
\end{equation}
where $\mathcal{E}_k^{\text{tail}}$ and $\mathcal{E}_k^{\text{head}}$ denote the tail and the head of the edge $\mathcal{E}_k$.
$|\mathcal{V}|$ and $|\mathcal{E}|$ refer to the cardinality of the sets $\mathcal{V}$ and $\mathcal{E}$ respectively.

Each node in $\mathcal V$ can be associated to the agents' position vector ${p}=\text{col}({p}_1,\cdots ,{p}_{n})$ by indicating the $i$-th agent's position with $p_i\in\mathbb{R}^2$. 
Since for any $k$-th edge in $\mathcal E$, with pairing agents $(i,j)$, we can define the relative position vector between the agents $i$ and $j$ as ${z}_k={p}_i-{p}_j\in\mathbb{R}^2$. Then, every edge in $\mathcal E$ can be associated to the vector of relative positions ${z}=\text{col}({z}_1,\cdots,{z}_{|\mathcal{E}|})$.
The vector $z$ can be described, in compact form, by
\begin{equation*}
    {z}=\bar{{B}}^T{p}
\end{equation*}
with $\bar{B}={B}\otimes I_{2}\in \mathbb{R}^{2n\times 2|\mathcal{E}|}$.

Using the above graph formalism, we can formalize the notion of infinitesimally rigid formation as follows.
With the formation {\it framework} defined by the tuple $(\mathbb{G},{p})$, an {\it edge function} $f_{\mathbb{G}}:\mathbb{R}^{2n}\rightarrow \mathbb{R}^{|\mathcal{E}|}$ is defined by 
\begin{equation*}
    f_{\mathbb{G}}({p}):={\text{col}}(\lVert z_1\rVert^2,\cdots,\lVert z_{|\mathcal{E}|}\rVert^2)=D_{{z}}^\top
    {z}
\end{equation*}
where $D_{z}:=\text{diag}(z_k)\in\mathbb{R}^{2|\mathcal{E}|\times |\mathcal{E}|}$ for $k=1,\cdots,|\mathcal{E}|$ is a block diagonal matrix.
The {\it rigidity matrix} $R(z)$ of the framework $(\mathbb G,p)$ is given by the Jacobian of the edge function $f_{\mathbb G}$ $R(z)=D_{{z}}^T\bar{{B}}^T$. 

For the formation in 2D plane as pursued in this paper, the framework $(\mathbb G,p)$ is said to be {\it infinitesimally rigid} if $\text{Rank}(R({z}))=2n-3$. 
Moreover, if $|\mathcal{E}|=2n-3$ it is said to be {\it minimally rigid}.  
Roughly speaking, for an infinitesimally and minimally rigid framework, the only group motions that can be performed on the whole group when they are already in the desired formation shape, are translations and rotations \cite{de2016distributed}. 
In this work, we assume that the agents' position vector dynamics is described by
\begin{equation}
    \label{eq:kinpoint}
    \dot{p}(t)=u(t) 
\end{equation}
where $\dot{p}(t)=:\frac{\dd}{\dd t}p(t)$ and $u(t)=\text{col}({u}_1(t),\cdots,{u}_n(t)$) is the concatenated control input signal with $u_i(t)$ be the velocity control input signal of the agent $i$. In the rest of the paper, we do not write the dependence on time $t$ in all signals when it is clear from the context. 
Let
\begin{equation}
    \label{eq:distance}
    d=\text{col}\left(d_1,\cdots,d_{|\mathcal{E}|}\right)\in\mathbb{R}^{|\mathcal{E}|}
\end{equation}
be the set of desired inter-agent distances associated to the desired formation shape. 
Accordingly, we can define the set of all equilibrium points that satisfy the desired distance constraints by $\mathcal{D}:=\{{p}\in\mathbb{R}^{2n}:\lVert{z}_k\rVert=d_k,\ \forall\ k=1,\cdots,|\mathcal{E}| \}$.

One of the well-known distributed formation control law that can guarantee the local exponential stability of $\mathcal{D}$ is the distance-based formation control, as expounded in \cite{Oh2015survey,Marina-book,Marina2015controlling}. 
In particular, the distributed formation control law is given by 
\begin{equation}
    \label{eq:grad_law1}
    {u}=\underset{ i\in\{1,\cdots,n\}}{\text{col}}({u}_i)=-\bar{{B}}D_{{z}}D_{\tilde{z}}{e}=R({z})^TD_{\tilde{z}}{e}
\end{equation}
where $\tilde{z}={\text{col}}(\lVert{z}_1\rVert^{\ell-2},\cdots,\lVert{z}_{|\mathcal{E}|}\rVert^{\ell-2})$ and ${e}={\text{col}}\big((\lVert{z}_1\rVert^\ell-d_1^\ell),\cdots,(\lVert{z}_{|\mathcal{E}|}\rVert^\ell-d_{|\mathcal{E}|}^\ell)\big)$ with $\ell\in\mathbb{N}$ \cite{de2016distributed}. 
In the following, we will only consider the case with $\ell=2$.

Note that the formation control law in \eqref{eq:grad_law1} is written in the most compact form. 
Looking at the individual control input we see that the local control law uses only local information available to each agent $i$, since
\begin{equation}
\label{eq:grad_law2}
{u}_i=-\sum_{k\in\mathcal{N}_i}\bar{B}_{ik} z_k \lVert{z}_k\rVert^{\ell-2} e_{k} \enspace,
\end{equation}
where $\bar{B}_{ik}$ is the $(i,k)$ element of the matrix $\bar{B}$. 
In other words, \eqref{eq:grad_law2} is a distributed control law for any $i$-th agent. 
When $\ell=2$, \eqref{eq:grad_law2} becomes
\begin{equation}
    \label{eq:grad_law3}
    {u}_i=-\sum_{k\in\mathcal{N}_i}\bar{B}_{ik} z_k e_{k} \enspace,
\end{equation}
which we will consider throughout this paper. 

\section{Mixed Uniform-Logarithmic Quantizer}\label{sec:MULQ}   

While the use of distributed formation control as described in Subsection \ref{subsec:formation} gives rise to distributed input signals defined on $\rline$, it cannot directly be implemented in the cyphertext $\mathbb{Z}_q$ in the edge or cloud via FHE-LWE method as presented in Section \ref{subsec:FHE}. 
Here, we present a mixed uniform-logarithmic quantizer method that allows the quantization of the sensing information, which are subsequently used to compute the formation control input in cyphertext. 
Although, such quantizer has been used before in \cite{chaher2021homomorphic}, its focus was on the deployment of distributed estimator in the cloud to compensate for sensor bias without stability analysis. 
In contrast, by restricting to the standard distributed formation control law \eqref{eq:grad_law3}, we show that the use of this quantizer can still guarantee the local asymptotic stability of the desired formation shape in our main result in Section \ref{sec:formation_cloud_control}. 

More precisely, the {\it mixed uniform-logarithmic quantizer} (MULQ) operator $Q:\rline\to\rline$ is defined by 
\begin{equation}
    \label{eq:quant}
    Q(x) := \dfrac{1}{S(x)}\left \lceil S(x)x \right \rfloor 
\end{equation}
where $S(x)$ is the {\it base-$10$ scaling factor} defined by
\begin{equation}
    \label{eq:scaling}
    S(x)=10^{(\sigma-\left \lfloor\log ( |x|) \right \rfloor-1)}
\end{equation}
$\left \lfloor \cdot \right \rfloor$ refers to the floor function and $\sigma\geq 1 \in \mathbb{N}$ is the desired {\it significant figures} parameter, i.e. the number of leading digits of $x$ that will be kept after the quantization.
For example, for $\sigma=1$ the set of quantization levels is given by $\left\{a\rho:a\in\{-9,-8,\ldots,8,9\},\  \rho\in\{10^{k}\},\ k\in\mathbb{N} \right\}= \{\ldots,0.09,0.1,0.2,\ldots,0.9,1,2,\ldots, 9,10,20,\ldots\}$.
This is in contrast to that of the standard logarithmic quantizer with coarseness parameter $0<\rho<1$ given by $\{\pm\rho^k : k\in\nline\}$ \cite{fu2005sector}. 

The significant figures parameter $\sigma$ plays an important role in guaranteeing the stability of the closed-loop system that depends on the information on the formation rigidity matrix and graph, as shown in Section \ref{sec:formation_cloud_control}.
In the following lemma, we show that the sector bound property of the proposed MULQ operator, where the bounds depend directly on the parameter $\sigma$. 
The larger $\sigma$ is, the thinner the band of the sector bound is and the closer it is to an identity operator. \vspace{0.2cm}
\begin{lemma}\label{lem:sector_bound}
Let the MULQ $Q:\rline\to\rline$ and $S(x)$ with a given $\sigma\geq 1$
be defined according to \eqref{eq:quant} and \eqref{eq:scaling}.
Then the following inequalities hold for all $x\in\rline$:
\begin{description}
\item[{\bf (A1).}] 
$\left(1-\frac{0.5}{10^{\sigma-1}}\right)|x|^2 \leq xQ(x)\leq\left(1+\frac{0.5}{10^{\sigma-1}}\right)|x|^2$
\item[{\bf (A2).}] $\left|x-Q(x)\right|\leq \frac{0.5}{10^{\sigma-1}}|x|$
\item[{\bf (A3).}] $|Q(x)| \leq \left(1+\frac{0.5}{10^{\sigma-1}}\right)|x|$
\end{description}
\end{lemma}

\begin{proof}
In the following we prove the lemma for the case $x>0$ only, since it is similar for $x<0$.
We first prove {\bf (A1)} with $x>0$. 
The difference between $x$ and $Q(x)$ can be rewritten as:
\begin{align*}
x-Q(x) & = x - \frac{1}{S(x)}\lceil S(x)x \rfloor = \frac{1}{S(x)}\left(S(x)x-\lceil S(x)x \rfloor \right) \enspace .
\end{align*}
Note, that the rounding operation gives us $-0.5 \leq S(x)x-\lceil S(x)x \rfloor \leq 0.5$. 
Hence, using the upper-bound of the rounding operation, it follows that:
\begin{align*}
x-Q(x)  \leq \frac{0.5}{S(x)} 
\Leftrightarrow Q(x) &\geq x-\frac{0.5}{S(x)} = x-\frac{0.5}{10^{\sigma-1}}10^{\lfloor \log x\rfloor} \\
                     &\geq x-\frac{0.5}{10^{\sigma-1}}x  = \left(1-\frac{0.5}{10^{\sigma-1}}\right)x \enspace,
\end{align*}
where we have used the fact that $10^{\lfloor \log x\rfloor}<x$ for positive $x$. 
This implies that $xQ(x) \geq \left(1-\frac{0.5}{10^{\sigma-1}}\right)x^2$. 
Similarly, using the lower-bound of $-0.5 \leq S(x)x-\lceil S(x)x \rfloor$, it follows that:
\begin{align}
\nonumber x-Q(x)  \geq \frac{-0.5}{S(x)} 
\Leftrightarrow Q(x) 
&\leq x + \frac{0.5}{S(x)} = x+\frac{0.5}{10^{\sigma-1}}10^{\lfloor \log x\rfloor} \\
\label{eq:q_upperbound} 
&\leq x+\frac{0.5}{10^{\sigma-1}}x  = \left(1+\frac{0.5}{10^{\sigma-1}}\right)x \ .
\end{align}
In this case, the upper-bound of $xQ(x)$ is given by $xQ(x) \leq \left(1+\frac{0.5}{10^{\sigma-1}}\right)x^2$, which proves {\bf (A1)}. 
Now similarly, to prove {\bf (A2)} with $x>0$ we rewrite:
\begin{align*}
    x-Q(x) & \leq \frac{0.5}{S(x)}  = \frac{0.5}{10^{\sigma-1}}10^{\lfloor \log x\rfloor} \leq \frac{0.5}{10^{\sigma-1}}x \enspace ,
\end{align*}
which implies immediately that $|x-Q(x)| \leq \frac{0.5}{10^{\sigma-1}} |x|$. 
Finally, from \eqref{eq:q_upperbound} follows that $|Q(x)| \leq \left(1+\frac{0.5}{10^{\sigma-1}}\right) |x|$, which proves {\bf (A3)}.
\end{proof}

Using the MULQ operator $Q$ as above, we can now encrypt the digits of $Q(x)$ by using $Q(x)S(x)\in [a]$ with $a=2(10)^\sigma-1$ and $[a]$ be as defined in Subsection \ref{subsec:FHE}. 
The cyphertext message $\text{Enc}(Q(x)S(x))$ or $\text{Enc}2(Q(x)S(x))$ can be sent to a third-party computing facility for further computation in the cyphertext space. 
Recalling the previous example with $\sigma=1$, we have that $Q(x)S(x)\in [19]$ for any $x\in\rline$. Thus a static plaintext space with cardinality $a=19$ can be used. 
The processed information in cyphertext can then be decrypted with local secret keys and re-scaled back by $\frac{1}{S(x)}$ to get the desired local control law. 
Again, using the example with $\sigma=1$, $\frac{1}{S(x)}\in\{10^k : k\in\nline\}$. 
This will be discussed further in the following section. 

\section{Distance-based Secure Formation Control}
\label{sec:formation_cloud_control}

\subsection{Distributed distanced-based formation control via MULQ and FHE-LWE}

The reason to employ a third-party computing facility to calculate the control inputs is to facilitate the use of external sensor systems that are essential to the control task.
In this case FHE-LWE ensures that the private information, coming from these sensors, cannot be retrieved by others.
In the previous section, we briefly discussed how $Q(x)$ and the digit information $S(x)Q(x)$ can be used to deploy FHE-LWE.
For every edge $k$, we have to encrypt the scalars $Q(z_{k,1})S(z_{k,1})$, $Q(z_{k,2})S(z_{k,2})$ and the scalar $Q(e_{k})S(e_{k})$ to enable the use of FHE-LWE on relative position vector $z_k=\text{col}(z_{k,1},z_{k,2})$ and on distance error $e_k$ information. 
The first two are encrypted with Enc2($\cdot$) via \eqref{eq:Enc2}, while the last one with Enc($\cdot$) via \eqref{eq:Enc} 
\begin{align*}
    \mathbf{Z}_{k,1}^{\text{Enc2}}&=\text{Enc2}\left(Q(z_{k,1})S(z_{k,1})\right)\\
    \mathbf{Z}_{k,2}^{\text{Enc2}}&=\text{Enc2}\left(Q(z_{k,2})S(z_{k,2})\right)\\
    \mathbf{E}_{k}^{\text{Enc}}   &=\text{Enc}\left(Q(e_{k})S(e_{k})\right)
\end{align*}

The scaling information will also be transmitted encrypted or un-encrypted to the corresponding agents in edge $k$ so that the processed information can be re-scaled back. 
When $\mathbf{U}_{i,k}^{\text{Enc}}$ is the resulting gradient computation of formation control law in the cyphertext for agent $i$ in $k$-th edge, the applied local control law for agent $i$ is given by:
\begin{align}
\label{eq:ui_rescaling}
u_i & = -\sum_{k\in\mathcal{N}_i}\bbm{\frac{\bar{B}_{ik}}{S(z_{k,1})S(e_{k})} \text{Dec}\big(\mathbf{U}_{i,k}^{\text{Enc}}\big)_1  \\
\frac{\bar{B}_{ik}}{S(z_{k,2})S(e_{k})} \text{Dec}\big(\mathbf{U}_{i,k}^{\text{Enc}}\big)_2} \enspace,
\end{align}
where we have used the decryption process Dec($\cdot$) in \eqref{eq:Dec} and  
$\text{Dec}\big(\mathbf{U}_{i,k}^{\text{Enc}}\big)_j$ refers to the $j$-th element of the vector $\text{Dec}(\mathbf{U}_{i,k}^{\text{Enc}})$. 
Following the multiplication property of FHE-LWE as in \eqref{eq:mult2}, it follows from above that
\begin{align}
\nonumber    u_i & = -\sum_{k\in\mathcal{N}_i}\bbm{\frac{\bar{B}_{ik}}{S(z_{k,1})S(e_{k})} \text{Dec}\left(\mathbf{Z}_{k,1}^{\text{Enc2}}\circledast \mathbf{E}_{k}^{\text{Enc}}\right)\\
    \frac{\bar{B}_{ik}}{S(z_{k,2})S(e_{k})} \text{Dec}\left(\mathbf{Z}_{k,2}^{\text{Enc2}}\circledast \mathbf{E}_{k}^{\text{Enc}}\right)} \\
\nonumber    & = -\sum_{k\in\mathcal{N}_i}\bbm{\frac{\bar{B}_{ik}}{S(z_{k,1})S(e_{k})} Q(z_{k,1})S(z_{k,1})Q(e_{k})S(e_{k})  \\
    \frac{\bar{B}_{ik}}{S(z_{k,2})S(e_{k})} Q(z_{k,2})S(z_{k,2})Q(e_{k})S(e_{k}) } \\
\label{eq:FHE-LWE-quantized}    & = -\sum_{k\in\mathcal{N}_i}\bar{B}_{ik}Q(z_k)Q(e_k)
\end{align}
where we define $Q(z_k):=\text{col}(Q(z_{k,1}),Q(z_{k,2}))$. 

In comparison to the unencrypted version in \eqref{eq:grad_law3} the local control law above contains the quasi-logarithmic quantized version of $z_k$ and $e_k$. 
In particular, FHE-LWE in the feedback loop can simply be regarded as an identity operator. 
In other words, FHE-LWE becomes transparent due to the use of MULQ operator and the closed-loop system analysis becomes an absolute stability analysis  with quantizers in the feedback loop. 
Consequently, for the analysis of closed-loop systems in the following subsection, the compact form of the whole formation control input can be written as 
\begin{equation}
    \label{eq:compact_quantized}
    u = -\bar{B}D_{Q(z)}Q(e)
\end{equation}
where the MULQ operator $Q$ is understood element-wise. 

Let us remark on securing the information of the scaling factor $S(\cdot)$ for both $z_k$ and $e_k$.
In the discussion above, this information is transmitted directly to the agent and used to re-scale back the computed control input. This re-scaling operation can be secured in the following way. 
In addition to encrypting the quantized information of $Q(z_{k,j})S(z_k)$ and $Q(e_k)S(e_k)$, the sensing node can encrypt the exponent of $S(z_{k,j})$ and of $S(e_k)$, indicated here with $\mathbf{S}^{\text{Enc}}_{z,k,j}$ and $\mathbf{S}^{\text{Enc}}_{e,k}$ respectively, and send them to the remote computing facility.
The latter subsequently computes the addition operation of both $\mathbf{S}^{\text{Enc}}_{z,k,j}$ and $\mathbf{S}^{\text{Enc}}_{e,k}$ and the result is transmitted to the corresponding agents. 
The agent can then perform the re-scaling operation by using the fact that 
\begin{equation*}
S(z_{k,j})S(e_k) = 10^{ \text{Dec}\left(\mathbf{S}^{\text{Enc}}_{z,k,j} + \mathbf{S}^{\text{Enc}}_{e,k}\right)}
\end{equation*}
for dimensions $j=1,2$ for substitution in \eqref{eq:ui_rescaling}.

\subsection{Absolute stability analysis of the closed-loop systems}

The application of FHE-LWE \eqref{eq:FHE-LWE-quantized} to the formation control of \eqref{eq:grad_law1} using quantized values of $Q(z_{k,1})S(z_{k,1})$, $Q(z_{k,2})S(z_{k,2})$ and $Q(e_{k})S(e_{k})$ becomes equivalent to the ones obtained without FHE-LWE, which is compactly written in \eqref{eq:compact_quantized}. 
Correspondingly, in the following proposition, we will analyze the stability of the closed-loop system where the FHE-LWE operation is replaced by an identity operator. \vspace{0.2cm}

\begin{proposition}\label{prop1}
Consider the mobile robots whose dynamics are given by \eqref{eq:kinpoint}. 
Suppose that the control inputs are given by the distributed quantized gradient-based formation control law \eqref{eq:compact_quantized} with the desired formation shape defined by the desired distance vector $d$ as in \eqref{eq:distance} and the mixed uniform-logarithmic quantization operator $Q$ be as in \eqref{eq:quant} with significant figures constant $\sigma\geq 1$.
Assume that the formation graph is infinitesimally and minimally rigid and connected. 
Then for sufficiently large $\sigma$, the equilibrium point $e=0$ is locally asymptotically stable. 
\end{proposition}\vspace{0.2cm}

\begin{proof}
The proof is based on the established local asymptotic stability results in distance-based formation control and we refer interested reader to \cite{de2016distributed,Marina2015controlling,Marina-book,AnYuFiHe08} among many others. 
The dynamics of the closed-loop autonomous multi-agent system can be written as
\begin{align}
\label{eq:closed-loop_sys1}
&\dot{z}=\bar{B}^T\dot{p}=-\bar{B}^T\bar{B}D_{Q(z)}Q(e) \\
\label{eq:closed-loop_sys2}
&\dot{e}=D_z^T\dot{z}=-D_z^T\bar{B}^T\bar{B}D_{Q(z)}Q(e) \enspace,
\end{align}
where as before the MULQ operator $Q$ is understood element-wise when a vector is used in its argument. 

Let us consider the following standard Lyapunov function as used in the aforementioned papers
\begin{equation}
\label{eq:Ve}
V=\frac{1}{4}e^Te=\sum_{k=1}^{|\mathcal{E}|}V_k=\frac{1}{4}\sum_{k=1}^{|\mathcal{E}|}(\lVert z_k\rVert^2-d_k^2)^2 \enspace.
\end{equation}
By computing its time-derivative along the trajectory of the closed-loop systems, we have
\begin{align}
\nonumber\dot V = & -e^TD_z^T\bar{B}^T\bar{B}D_{Q(z)}Q(e) \\
\nonumber       = &+\frac{1}{2}\left(D_ze-D_{Q(z)}Q(e)\right)^T\bar{B}^T\bar{B}\left(D_ze-D_{Q(z)}Q(e)\right) \\
\label{eq:V_dot}  & - \frac{1}{2}e^TD_z^T\bar{B}^T\bar{B}D_ze - \frac{1}{2}Q(e)^TD_{Q(z)}^T\bar{B}^T\bar{B}D_{Q(z)}Q(e) \enspace,
\end{align}
where $\bar{B}$ describes the incidence matrix of 
formation graph $\mathbb{G}$. 
In this case, $\bar{B}^T\bar{B}$ is positive semi-definite matrix with the kernel being a vector of ones $\mathds{1}$, due to the connectedness of the undirected graph. 

As established in literature of distance-based formation control (c.f. \cite{de2016distributed,Marina2015controlling,Marina-book,AnYuFiHe08}), the second term on the right-hand side is negative definite and satisfies
\begin{equation}
\label{eq:second_term}
- \frac{1}{2}e^TD_z^T\bar{B}^T\bar{B}D_ze \leq - \lambda_{\min}\lVert e\rVert^2
\end{equation}
where $\lambda_{\min}$ refers to the smallest eigenvalue of the positive definite matrix $D_z^T\bar{B}\bar{B}^TD_z$ in the neighborhood of $e=0$. 

The last term of \eqref{eq:V_dot} is upper-bounded by zero as $\bar{B}\bar{B}^T$ is positive semi-definite. We will now compute the upper-bound of the first term in \eqref{eq:V_dot} as follows
\begin{align}
\nonumber & \frac{1}{2}\left(D_ze-D_{Q(z)}Q(e)\right)^T\bar{B}^T\bar{B}\left(D_ze-D_{Q(z)}Q(e)\right) \\
\nonumber & = \frac{1}{2}\lVert\bar{B}\left(D_ze-D_{Q(z)}Q(e)\right)\rVert^2\\
\nonumber & \leq \frac{1}{2}\lVert\bar{B}\left(D_ze-D_zQ(e)\right)\rVert^2 + \frac{1}{2}\lVert\bar{B}\left(D_zQ(e)-D_{Q(z)}Q(e)\right)\rVert^2 \\
\label{eq:first_term}& = \frac{1}{2}\lVert\bar{B}D_z\left(e-Q(e)\right)\rVert^2 + \frac{1}{2}\lVert\bar{B}\left(D_z-D_{Q(z)}\right)Q(e)\rVert^2 \enspace.
\end{align}
By Lemma \ref{lem:sector_bound} the three inequalities:
$\rVert e-Q(e)\lVert  \leq \frac{0.5}{10^{\sigma-1}}\lVert e\rVert$, 
$\lVert D_z-D_{Q(z)}\rVert  \leq \frac{0.5}{10^{\sigma-1}}\lVert z\rVert$, and 
$\lVert Q(e) \rVert  \leq \left(1+\frac{0.5}{10^{\sigma-1}}\right)\lVert e\rVert$ hold.
Combining these terms to \eqref{eq:first_term} and together with \eqref{eq:second_term}, it follows that \eqref{eq:V_dot} becomes
\begin{align*}
\dot V 
\leq &\lambda_{\max}\left(\frac{0.5}{10^{\sigma-1}}+\frac{0.5}{10^{\sigma-1}}
                    \left(1+\frac{0.5}{10^{\sigma-1}}\right)\right)
\lVert z\rVert^2\lVert e\rVert^2 \\& 
- \lambda_{\min}\lVert e\rVert^2
\end{align*}
where $\lambda_{\max}>0$ is the maximum eigenvalue of $\bar{B}^T\bar{B}$. 

Note that $\lVert z\rVert$ can be expressed as a continuous function of $e$, namely, $\lVert z\rVert=\sqrt{\sum_k |e_k+d_k^2|}$. 
Thus in the neighborhood of $e=0$, e.g. in  
$\mathbb{B}_\delta:=\{e:\lVert e\rVert<\delta\}$, $\lVert z\rVert^2$ can be upper bounded by a constant $c$ that depends on the desired distance $d$ and the radius of the neighborhood $\delta$. 
Correspondingly, for a sufficiently large $\sigma$, the right-hand side of the above inequality can be made negative in $\mathbb{B}_\delta$ such that 
\begin{equation}
\label{eq:V_dot_k}
\dot V \leq - k\lVert e\rVert^2
\end{equation}
for all $e\in \mathbb{B}_\delta$ with $0<k<\lambda_{\min}$ and in particular
\begin{align*}
k = &\lambda_{\min} - \lambda_{\max}c\left(\frac{0.5}{10^{\sigma-1}}+\frac{0.5}{10^{\sigma-1}}\left(1+\frac{0.5}{10^{\sigma-1}}\right)\right) \enspace.
\end{align*}
This implies that $\mathbb{B}_\delta$ is forward invariant, so that $\lVert z(t)\rVert$ is bounded by $c$ for all $t\geq 0$ and $\lVert e(t)\rVert\to 0$ as $t\to\infty$. 
In other words, the formation converges to the desired shape. 
\end{proof}

We note that a different value of $\sigma$ can be assigned in the quantization of $z_{k,j}$ and of $e_k$ in order to get a trade-off between asymptotic stability and minimizing the required plaintext space. 
On the one hand, as shown in the proof of Proposition \ref{prop1}, the parameter corresponding to $e$ (denoted conveniently as $\sigma_{e_k}$) plays a crucial role in ensuring that \eqref{eq:V_dot_k} holds. 
It has to be chosen sufficiently large for asymptotic stability. 
On the other hand, the parameter for $z_{k,j}$ (denoted as $\sigma_{z_{k,j}}$) can be assigned to $1$ safely. 
It allows us to minimize the space of plaintext needed for the encryption and decryption.

\section{Numerical Simulation}
\label{sec:numerical_sim}

In this section, we present the results of a numerical simulation implemented in Python. Similar to \cite{chaher2021homomorphic} we used Python to circumvent the integer overflow problem by employing arbitrary-precision integers. 
The task of the distance-based secure edge controller is to guide a system of 4 agents into a square formation with desired inter-agent distances. 
The undirected graph $\mathbb{G}=(\mathcal{V},\mathcal{E})$ is defined with agents $\mathcal{V}=\{1,2,3,4\}$ and $\mathcal{E}=\{(1,2),(2,3),(1,3),(3,4),(1,4)\}$, while 
the desired distance vector is $d=[1\ 1\ \sqrt{2}\ 1\ 1]^T\in\mathbb{R}^5$.
The initial conditions for the agents' position vector $p(0)$ is randomly generated within the basin of attraction.
Regarding the encryption, the plaintext $[a]$ and cyphertext $\mathbb{Z}_q$ space available are defined by $a=10^{11}$ and $q=10^{22}$.
The secret key vector $\mathbf{s}$ length is $N=30$, while the sampling space $[r]$ of the injected error vector $\mathbf{e}$ is defined by $r=4$ so that the conditions for \eqref{eq:HE+} and \eqref{eq:mult2} hold. 
Following the computation in the proof of Proposition 3.1 and using the neighborhood $\mathbb{B}_{2.7}$ of $e=0$, we can obtain the constants $c\approx 12.04$, $\lambda_{\text{min}}\approx 0.058$ and $\lambda_{\text{max}}\approx 4.11$. By taking $\sigma = 4$, the condition stated after \eqref{eq:V_dot_k} in the proof of Proposition \ref{prop1} is satisfied. 

\begin{figure}[t]
\centering
\begin{subfigure}{.5\columnwidth}
  \centering
  \includegraphics[width=1\columnwidth,clip,trim={0.4cm 0.5cm 0 0}]{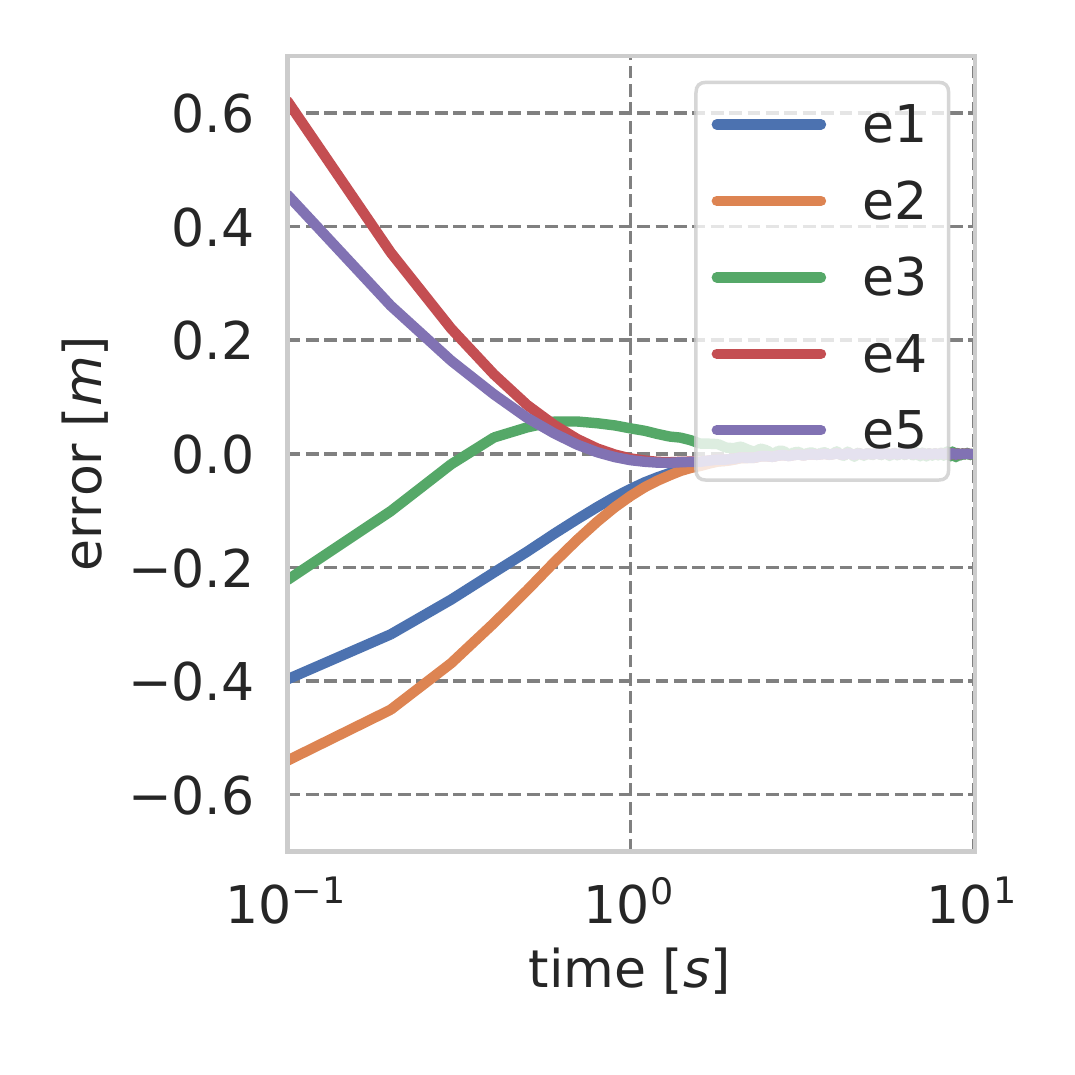}
  \caption{Error components $e_i(t)$.}
  \label{fig:sub1}
\end{subfigure}%
\begin{subfigure}{.5\columnwidth}
  \centering
\includegraphics[width=1\columnwidth,clip,trim={0.4cm 0.5cm 0 0}]{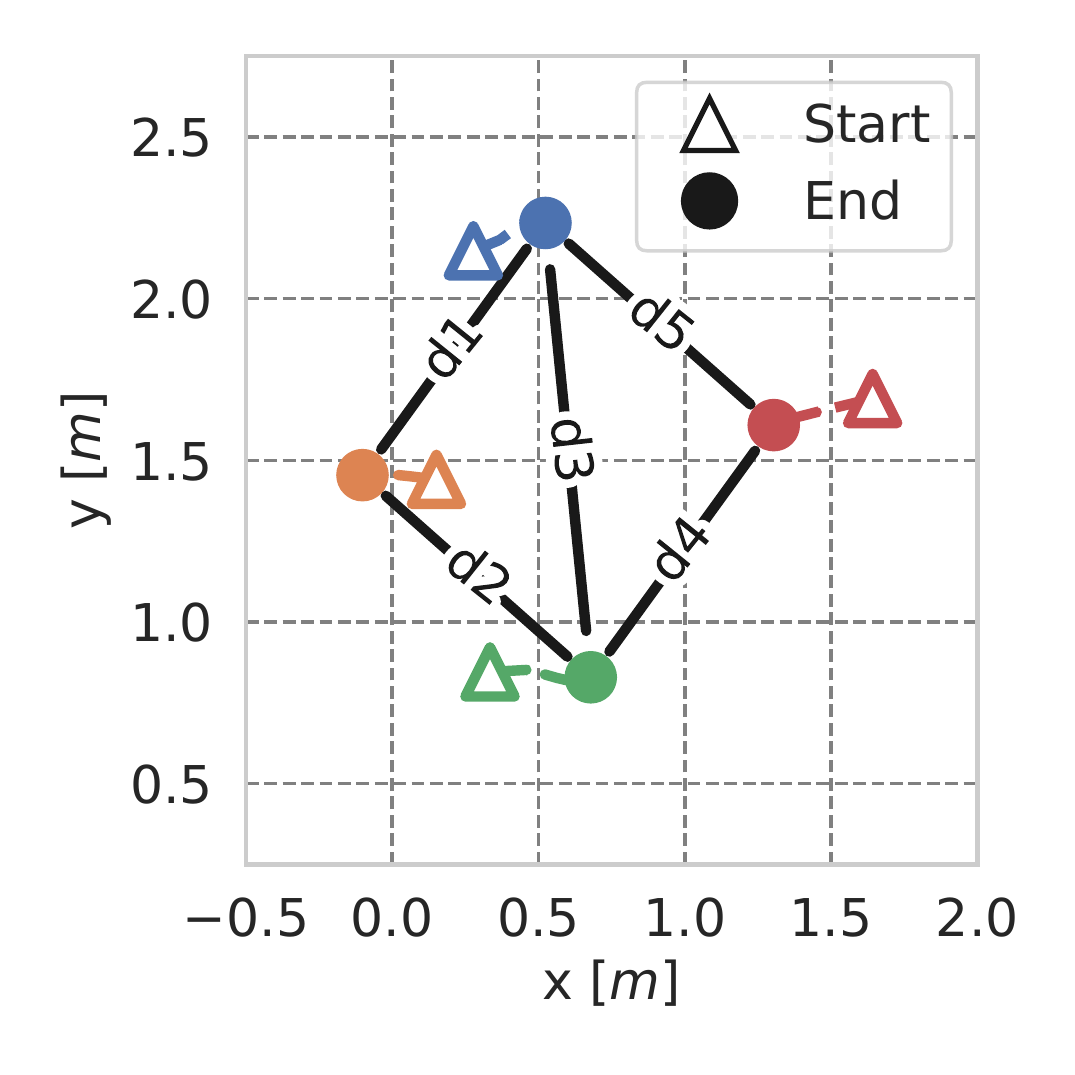}
\caption{Agent trajectories.}
  \label{fig:sub2}
\end{subfigure}
\caption{Secure formation control simulation with four agents forming a square: 
(a) error trajectories over time (in semi-logarithmic scale); 
(b) top view of the agents' 2D trajectories of the four agents starting from the initial conditions (triangles) and converging towards the desired square formation (filled circles).}
\label{fig:simulation}
\end{figure}

Using the above simulation setup, the corresponding simulation result is shown in Figure \ref{fig:simulation}. 
Figure \ref{fig:simulation}a) shows the plot of error signal $e_i$ from all five edges $i=1,\ldots 5$. 
It demonstrates that the error vector $e\in\mathbb{R}^5$ of the multi-agent system converges to the equilibrium point $e=0$ as expected with an exponential rate of convergence. 
Panel \ref{fig:simulation}b) presents a top view of the agents' position vector $p\in\mathbb{R}^8$ over time. Each agent starts from its initial position depicted in triangle shapes and all agents converge exponentially to the desired shape of a square (shown as filled circles in the figure). 

We remark here that the computation of  \eqref{eq:V_dot_k} leads to a conservative bound of the parameters. 
Indeed, in simulations, we can assign smaller values of $\sigma$ or larger values of $\delta$ than the ones given above for which the formation goal is still attained. 
\vspace{0.3cm}
\newpage
\section{Conclusion}
\label{conclusions}

In this paper, we proposed a secure distributed formation control system enabled by FHE-LWE encryption and MULQ quantization. 
While a similar framework has been presented before \cite{chaher2021homomorphic} with an empirical analysis, in this contribution we present rigorous analysis of the closed-loop systems.
Specifically, we show the sector bound property of the proposed MULQ and we present an absolute stability analysis showing the asymptotic stability of the closed-loop secure control system. Since we have shown that MULQ can be used together with FHE in the design of secure  formation control, the combined use of MULQ with FHE can be explored further in other secure control design problems.    

\bibliographystyle{IEEEtran}

\bibliography{root.bib}

\end{document}